# Efecto de la seguridad social en la duración del ausentismo laboral en el Servicio de Salud de Ñuble: un análisis de supervivencia

ARIEL SOTO CARO, ROBERTO HERRERA COFRÉ[a], RODRIGO FUENTES SOLÍS[a]

## Social security and labor absenteeism in a regional health service

Escuela de Administración y Negocios, Universidad de Concepción, Chillán, Chile.
[a]Economista, M.Sc.

***Background***: *Absenteism can generate important economic costs.* ***Aim***: *To analyze the determinants of the time off work for sick leaves granted to workers of a regional health service.* ***Material and Methods***: *Information about 2033 individuals, working at a health service, that were granted at least one sick leave during 2012, was analyzed. Personal identification was censored. Special emphasis was given to the type of health insurance system of the workers (public or private).* ***Results***: *Workers ascribed to the Chilean public health insurance system (FONASA) had 11 days more off work than their counterparts ascribed to private health insurance systems. A higher amount of time off work was observed among older subjects and women.* ***Conclusions***: *Age, gender and the type of health insurance system influence the number of day off work due to sick leaves.*
*(Rev Med Chile 2015; 143: 987-994).*

Sin desconocer el derecho a la salud que posee todo trabajador y el valor del reposo como indicación médica, el uso de este último implica dos tipos de costos para la sociedad. El primero se refiere a la disminución de la productividad, y el segundo dice relación con el subsidio que recibe el trabajador mientras se ausenta a sus labores, llamado Subsidio por Incapacidad Laboral (SIL). El SIL representa una parte importante[1] del costo por ausentismo laboral.

Las licencias se pueden dividir en enfermedad común o accidente, licencias por reposo maternal, por enfermedad grave del niño menor de un año y accidente laboral o de trayecto. El objetivo de las licencias del primer tipo es garantizar el reposo como indicación médica y representa la mayor concentración, tanto en fondos gastados como en cantidad de licencias y días de licencia[2], siendo los trastornos psiquiátricos la mayor patología de este ítem, y la que presenta tasas de crecimiento anuales considerables[3].

El costo de los subsidios por licencias médicas está directamente relacionado con la cantidad de días que duran las licencias. Para el año 2012, las Instituciones de Salud Previsional (ISAPRE) financiaron en 148.549 millones de pesos a trabajadores del sector privado sólo en licencias curativas, lo que representa 57,5% de todas las licencias, mientras que se financió en 46.589 millones de pesos a trabajadores del sector público (66,1% del total de licencias). Por otro lado, el Fondo Nacional de Salud (FONASA) financió en 300.758 millones de pesos a trabajadores del sector privado, también en licencias curativas (74,3% del total), y 36.039 millones para trabajadores del sector público. El aumento de estos costos marca 92,5% desde el 2001 al 2013[2], con aproximadamente una tasa de 9% anual[4], motivando preguntas sobre el correcto

o eficiente funcionamiento de este mecanismo de protección social. Algunos investigadores han demostrado el efecto del género y la edad como determinante de la cantidad de licencias médicas y su duración[3]. Aunque existe escasa evidencia sobre un posible abuso de las licencias para casos específicos[5], otros han defendido la idea que la percepción de tal abuso ha llevado a restringir la utilización de este instrumento para la prevención de la enfermedad y su uso como corrector de variables psicosociales, remitiéndose únicamente para efectos curativos de patologías declaradas[6].

Adicionalmente, la investigación de Ziebarth y Karlsson[7] muestra que en los países pertenecientes a la Organización para la Cooperación y el Desarrollo Económicos (OCDE) existe una amplia dispersión de los días medios de licencias, que van desde 4 a 29. Observando que existen distintos costos asociados para el trabajador, distintos copagos y distintas fiscalizaciones, de acuerdo al sistema de protección al que el trabajador se encuentre afiliado y considerando que el sector hospitalario presenta una alta tasa relativa de ausentismo[8], el propósito de esta investigación es revisar el efecto del tipo de protección social como determinante del riesgo de recibir una licencia médica curativa y la probabilidad de que ésta finalice en un período de tiempo dado. En términos biométricos, se realiza un análisis de supervivencia que estime, sobre los funcionarios que se ausentan, el efecto del sistema de protección social como determinante de la duración de la licencia y de los días trabajados. El estudio se realizó en el Servicio de Salud de Ñuble (SSÑ), ubicado en la región del Biobío, en Chile, conformado por un equipo de alrededor de 3.100 personas y otras 1.500 que se suman desde los Centros de Salud de atención primaria. En el año 2013 el SSÑ atendió más de 1,4 millones de consultas médicas, para una población que sufre principalmente de enfermedades circulatorias, según el Subdepartamento de Bioestadística e Información de Salud.

## Materiales y Métodos

### Diseño
Estudio retrospectivo, descriptivo, analítico y causal.

### Muestra
Acogiéndose a la Ley 20.285 sobre Transparencia y Acceso a la Información Pública, se solicitó al SSÑ una base de datos con 2.033 observaciones de individuos con identidad censurada, que han recibido, a lo menos, una licencia curativa durante el año 2012. Esta última situación hace imposible calcular algunas de las tasas definidas comúnmente en la literatura, por ejemplos: tasa de intensidad de uso, tasa de incapacidad laboral, etc.[5]. En consecuencia, se propone una "tasa de ausentismo censurada", la cual se define como el cociente entre los días de ausentismo laboral y los 365 días del año y luego multiplicado por 100 para interpretarlo como una tasa porcentual. Representa el porcentaje de ausentismo en el año para cada funcionario. Debido a que el interés de esta investigación radica en los determinantes de la duración de las licencias médicas, se justificó utilizar únicamente funcionarios que han sufrido ausentismo.

### Estadística descriptiva
En la Tabla 1 se presentan algunas características de la base de datos utilizada. El 27% son hombres y 73% son mujeres, 73% de los encuestados pertenece a FONASA y el restante a ISAPRE. Con respecto a las profesiones, 38% son Técnicos, 24% son Profesionales y 13% son Médicos. En relación a los establecimientos, 49% de los trabajadores se concentran en el Hospital Herminda Martín, seguido de 15% en el Hospital de San Carlos y 9% del SSÑ.

### Método biométrico
Como paso inicial se realizó un análisis ANOVA, donde la variable dependiente representa los días de ausentismo laboral, contra variables categóricas *individuales* (género, rango de edad y sistema de salud) e *institucionales* (planta, establecimiento y tipo de contrato).

La duración de las licencias fue analizada biométricamente a través de un modelo de supervivencia, tal como recomienda la Dirección de Presupuestos del Gobierno de Chile[4]. El fenómeno de duración fue revisado desde dos perspectivas: 1) Determinantes del riesgo (estadístico) de recibir una licencia médica, es decir, a medida que avanza el tiempo, cómo cambia la probabilidad de caer en ausentismo; 2) Determinantes del riesgo de la duración de dicha licencia, es decir, a medida que avanza el tiempo, cuál es el riesgo de que dicha licencia se prolongue.

Específicamente se utilizó un modelo de regre-sión semi-paramétrico de Cox[9,10]. Dicho modelo permite estimar el riesgo de que los individuos experimenten un evento específico, medido sobre el tiempo. Para estudiar el riesgo de recibir una licencia, se consideró como variable dependiente la cantidad de días trabajados antes de requerir una licencia. Luego, para estimar los determinantes del riesgo de la duración de una licencia se utilizó como variable dependiente la cantidad de días que un trabajador estuvo bajo el régimen de una licencia médica antes de volver al trabajo.

**Tabla 1. Descripción de la base de datos del Servicio de Salud Ñuble**

| Variable | Categoría | % de la muestra | Rango de Edad (años) en % | | | | |
|---|---|---|---|---|---|---|---|
| | | | Menor a 31 | 31 a 40 | 41 a 50 | 51 a 60 | Más de 61 |
| Género | Masculino | 27 | 71 | 74 | 74 | 76 | 54 |
| | Femenino | 73 | 29 | 26 | 26 | 24 | 46 |
| Sistema de Salud | Fonasa | 73 | 66 | 72 | 76 | 77 | 73 |
| | Isapre | 27 | 34 | 28 | 24 | 23 | 27 |
| Planta | Administrativos | 11 | 8 | 10 | 16 | 12 | 10 |
| | Auxiliares | 11 | 3 | 6 | 15 | 19 | 17 |
| | Directivos | 1 | 0 | 0 | 0 | 2 | 3 |
| | Médicos | 13 | 17 | 12 | 12 | 9 | 19 |
| | Odontólogos | 3 | 4 | 3 | 2 | 2 | 1 |
| | Profesionales | 24 | 26 | 28 | 18 | 22 | 27 |
| | Técnicos | 38 | 42 | 42 | 37 | 34 | 24 |
| Establecimiento | Cesfam Violeta Parra | 5 | 6 | 5 | 7 | 4 | 5 |
| | Dir. Servicio de Salud Ñuble | 9 | 13 | 10 | 8 | 6 | 3 |
| | Hospital Clínico Herminda Martín | 49 | 40 | 46 | 51 | 58 | 56 |
| | Hospital de Bulnes | 5 | 6 | 4 | 5 | 6 | 5 |
| | Hospital de Coelemu | 5 | 7 | 6 | 4 | 3 | 6 |
| | Hospital de El Carmen | 4 | 3 | 4 | 4 | 3 | 1 |
| | Hospital de Quirihue | 5 | 7 | 4 | 5 | 4 | 8 |
| | Hospital de San Carlos | 15 | 15 | 18 | 12 | 14 | 13 |
| | Hospital de Yungay | 4 | 4 | 3 | 5 | 3 | 3 |
| Días de ausentismo (promedio) | | 23 | 15 | 20 | 21 | 30 | 43 |
| Total (n) | | 2.033 | 388 | 578 | 461 | 450 | 156 |

La función de riesgo utilizada se construyó como:

$$h_i(t) = h_0(t)e^{\beta X_i} \quad (1)$$

El riesgo base es $h_0$, que se vincula al tiempo que dura la licencia (o la suma de días donde un sujeto $i$ recibe una, o más de una licencia, durante un año, $\beta$ representa el efecto de las covariables ($X$) individuales e institucionales, las cuales son interpretadas como tasas de riesgo. El método muestra cómo las covariables aumentaron o disminuyeron el riesgo de ocurrencia de un determinado evento.

**Resultados**

Analizando la relación entre los días de ausentismo y las variables individuales mediante ANOVA (Tabla 2) se observó que las mujeres, en promedio, superaron en 6 días el ausentismo de los hombres, la media de la duración de las licencias médicas de las mujeres varió desde 22,6 a 26,9 días, mientras que el rango de los hombres varió desde 15,9 a 21,5 días. Los promedios de ausentismo por grupo de edad difirieron significativamente entre ellos, y se observó una relación positiva entre estas dos variables.

Además, al examinar la relación de los días de ausentismo y el sistema de salud se observó que las personas que pertenecen a FONASA tuvieron en promedio un ausentismo de 26 días, el cuál superó en 11 días a los que pertenecen a ISAPRE. Este fe-

**Tabla 2. Análisis de Varianza de Variables Individuales**

| Variables individuales | Categoría | Número de individuos (n) | Días de ausentismo (media) | Días de ausentismo (mediana) |
|---|---|---|---|---|
| Género* | Masculino | 557 | 19 | 8 |
| | Femenino | 1.476 | 25 | 12 |
| Rango de edad* | Menor a 31 | 388 | 15 | 8 |
| | 31 a 40 | 578 | 20 | 11 |
| | 41 a 50 | 461 | 21 | 10 |
| | 51 a 60** | 450 | 30 | 13 |
| | 61 y más** | 156 | 43 | 19 |
| Sistema de salud* | Fonasa | 1.481 | 26 | 12 |
| | Isapre | 552 | 15 | 8 |

*Variables significativas al 95% de confianza mediante Análisis de Varianza (ANOVA). **Categorías significativas mediante el test a *posteriori* de "T3 de Dunnett".

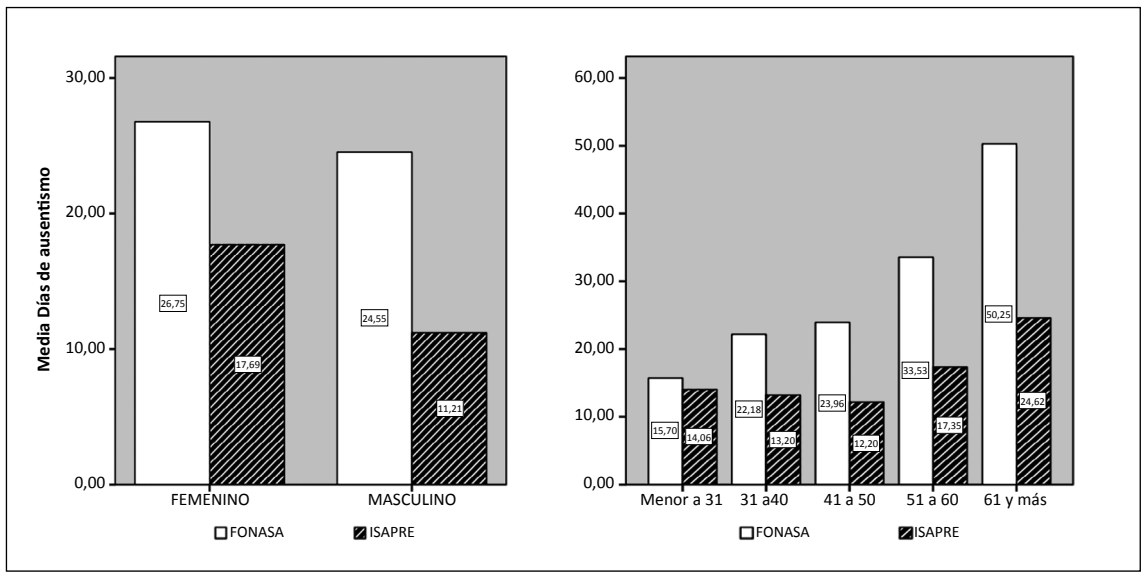

**Figura 1.** Días de ausentismo por sistema de salud según género y rango de edad.

nómeno fue transversal, independiente del género y del sistema de salud del trabajador (Figura 1).

La Tabla 3 señala que los trabajadores que pertenecen a las plantas de "Auxiliares" y "Técnicos" presentaron los mayores promedios de ausentismo laboral (30 y 27 días respectivamente), siendo significativamente opuestos a los de Médicos y Odontólogos (11 y 17 días respectivamente). Con respecto al tipo de contrato, los trabajadores que poseen la categoría de "Titular" presentaron los mayores promedios de ausentismo laboral (29 días), significativamente superior a los 20 días de licencia médica de los "Contratados".

En la etapa biométrica se estimaron dos modelos de supervivencia, utilizando como covariables las variables individuales (Tabla 2) e institucionales (Tabla 3). El primer modelo estima el riesgo de requerir una licencia, por lo que su variable dependiente es la cantidad de días de trabajo efectivo (sin licencia) que ha tenido un funcionario dentro de un año. El segundo modelo estima el riesgo de que la licencia se alargue.

Tabla 3. Análisis de Varianza de Variables Institucionales

| Variables institucionales | Categoría | Número de individuos (n) | Días de ausentismo (media) | Días de ausentismo (mediana) |
|---|---|---|---|---|
| Planta* | Administrativos | 229 | 25 | 11 |
| | Auxiliares | 222 | 30 | 14 |
| | Directivos | 13 | 23 | 5 |
| | Médicos** | 259 | 11 | 6 |
| | Odontólogos | 56 | 17 | 11 |
| | Profesionales | 484 | 21 | 11 |
| | Técnicos | 761 | 27 | 13 |
| Establecimiento* | Cesfam Violeta Parra | 107 | 18 | 8 |
| | Dir. Servicio de Salud Ñuble | 175 | 17 | 10 |
| | Hospital Clínico Herminda Martín | 1.005 | 22 | 11 |
| | Hospital de Bulnes | 97 | 32 | 11 |
| | Hospital de Coelemu** | 101 | 42 | 11 |
| | Hospital de El Carmen | 73 | 16 | 17 |
| | Hospital de Quirihue | 102 | 25 | 14 |
| | Hospital de San Carlos | 298 | 23 | 11 |
| | Hospital de Yungay | 75 | 25 | 12 |
| Tipo de contrato* | Contratado | 1.243 | 20 | 10 |
| | Titular | 790 | 29 | 13 |

*Variables significativas al 95% de confianza mediante Análisis de Varianza (ANOVA). **Categorías significativas mediante el test a *posteriori* de "T3 de Dunnett".

Aquí la variable dependiente es la cantidad de días, dentro de un año, que un funcionario está ausente de su trabajo. Este último modelo permite obtener los determinantes de la duración de una licencia médica. La Tabla 4 presenta los resultados de dichas estimaciones.

El género fue importante para determinar el riesgo de caer en ausentismo y la duración de la licencia. Los hombres tuvieron 12,3% menos riesgo de recibir una licencia y 14,8% de que ésta terminara antes que la de una mujer.

Por otro lado, a medida que las personas poseen más edad tuvieron mayor probabilidad de ausentarse a su trabajo, y una vez con licencia, también tuvieron mayor probabilidad de que sus licencias fueran más largas.

Otro resultado importante fue que, aunque el sistema de salud por sí solo no es determinante del riesgo de recibir una licencia médica, sí lo fue para determinar la duración de la licencia.

Además, considerando simultáneamente la edad y la cobertura sanitaria, las personas más jóvenes se ausentaron menos que las mayores, pero siempre quienes están en FONASA tuvieron más días ausentes que quiénes están en ISAPRE. Este efecto se mantuvo en la duración de la licencia. Es decir, los funcionarios mayores tuvieron licencias más largas, pero quienes están en ISAPRE se ausentaron menos que sus homólogos en FONASA.

El género fue significativo por sí solo, pero al interactuar con el sistema de salud, no existieron diferencias significativas, lo que significa que una mujer cotizante en una ISAPRE, versus una mujer cotizante en FONASA (con todas sus otras características iguales), tuvo el mismo riesgo de caer en ausentismo y sus licencias duraron lo mismo.

Dentro de todas las plantas o estamentos de trabajo, los "médicos" y los "directivos" tuvieron una diferencia significativa con respecto al riesgo de caer en ausentismo, observando que tuvieron 23,8% y 35,4% menos riesgo de recibir una licencia médica respectivamente, y una vez que se posee licencia, tanto los "médicos" como los "técnicos" tuvieron diferencias significativas con respecto del resto. Los médicos tuvieron 41,5%

**Tabla 4. Tasas de riesgo estimados para los análisis de regresión**

| Variables | | Riesgo requerir una licencia | | Riesgo de salir de la licencia | |
|---|---|---|---|---|---|
| Edad: rango (31-40) | | 1,2080*** | (0,0848) | 0,7677*** | (0,0602) |
| Edad: rango (41-50) | | 1,1461* | (0,0899) | 0,8181** | (0,0756) |
| Edad: rango (51-60) | | 1,3894*** | (0,1242) | 0,6227*** | (0,0606) |
| Edad: rango (>60) | | 2,1689*** | (0,2600) | 0,4347*** | (0,0486) |
| Salud: (Isapre =1, Fonasa = 0) | | 1,1960 | (0,1376) | 0,7351** | (0,0992) |
| Interacción | Edad: (31-40)/Isapre | 0,8454 | (0,1041) | 1,3520** | (0,1899) |
| | Edad: (41-50)/Isapre | 0,7728** | (0,0996) | 1,5015** | (0,2370) |
| | Edad: (51-60)/Isapre | 0,7598** | (0,1046) | 1,5682*** | (0,2470) |
| | Edad: (> 60)/Isapre | 0,5635*** | (0,1134) | 1,6983** | (0,3528) |
| Contrato: (Titular =1, Contratado = 0) | | 1,0417 | (0,0567) | 0,9315 | (0,0540) |
| Género: (Hombre = 1, Mujer = 0) | | 0,8772** | (0,0578) | 1,1475** | (0,0782) |
| Interacción Salud: Isapre/Género: Hombre | | 0,9389 | (0,0937) | 1,1369 | (0,1314) |
| Establecimiento: SSÑuble | | 1,2515** | (0,01424) | 0,7699* | (0,0978) |
| Establecimiento: Hospital HM | | 1,2482** | (0,1131) | 0,8331* | (0,0903) |
| Establecimiento: Bulnes | | 1,1372 | (0,1494) | 0,7361* | (0,1213) |
| Establecimiento: Coelemu | | 1,8376*** | (0,2441) | 0,5611*** | (0,0783) |
| Establecimiento: El Carmen | | 1,0523 | (0,1278) | 1,0159 | (0,1712) |
| Establecimiento: Quirihue | | 1,5383*** | (0,2025) | 0,7289** | (0,0968) |
| Establecimiento: San Carlos | | 1,3069*** | (0,1320) | 0,8144* | (0,0969) |
| Establecimiento: Yungay | | 1,3911*** | (0,1973) | 0,7437* | (0,1158) |
| Planta: Auxiliares | | 1,0397 | (0,1000) | 0,9107 | (0,0875) |
| Planta: Directivos | | 0,6462* | (0,1601) | 1,1696 | (0,4541) |
| Planta: Médicos | | 0,7621** | (0,0832) | 1,4145** | (0,1729) |
| Planta: Odontólogos | | 1,0925 | (0,1599) | 1,0865 | (0,1741) |
| Planta: Profesionales | | 0,9776 | (0,0811) | 1,0610 | (0,0959) |
| Planta: Técnicos | | 1,1260* | (0,0810) | 0,8730* | (0,0677) |
| Bondad de Ajuste: LR $\chi^2$ | | $LR\ \chi^2 = 148,92$ | | $LR\ \chi^2 = 199,69$ | |
| n | | 2.024 | | 2.024 | |

Donde *: valor-p < 0,1; **: valor-p < 0,05; ***: valor-p < 0,01. Las desviaciones estándar en paréntesis.

de riesgo de que su licencia terminara antes que el resto y los técnicos tuvieron 12,7% de riesgo de que su licencia terminara después del resto de los estamentos.

Las Figuras 2 y 3 presentan las curvas de supervivencia estimadas para cuatro perfiles de funcionarios. Mostrando, por ejemplo, que un hombre más joven en ISAPRE tuvo menos riesgo de recibir una licencia y tuvo licencias más cortas que un hombre mayor y en FONASA.

**Discusión**

Hemos comprobado algunos supuestos sobre el comportamiento del ausentismo laboral[3]. El género y la edad son variables influyentes para la solicitud y duración de las licencias médicas curativas. Las mujeres siempre se ausentan más que los hombres, sin importar su cobertura. Es interesante que el riesgo aumenta con la edad, por lo que podría inferirse que realmente las licencias están asociadas a indicaciones médicas pertinentes debido a que, a mayor edad, hay mayor prevalencia de enfermedades.

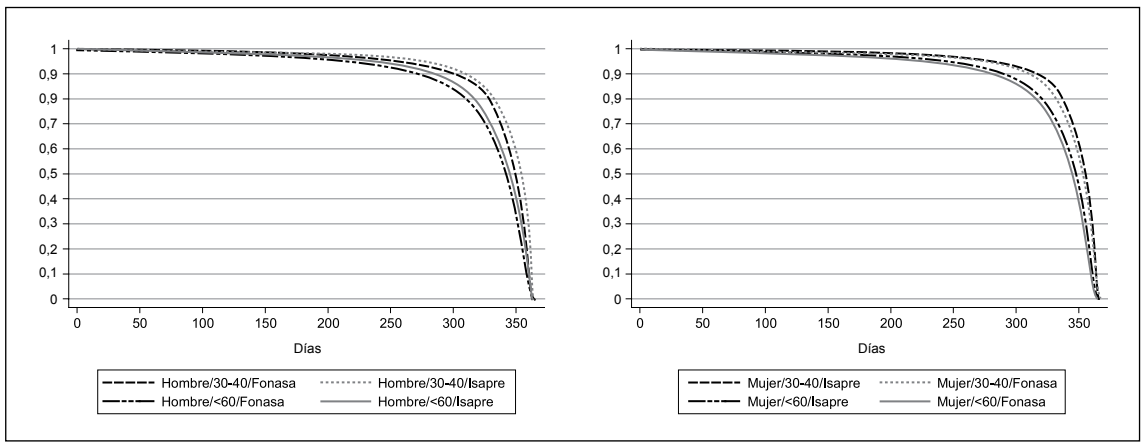

**Figura 2.** Curvas de supervivencia estimada por género, edad y sistema de salud, para el riesgo de recibir una licencia.

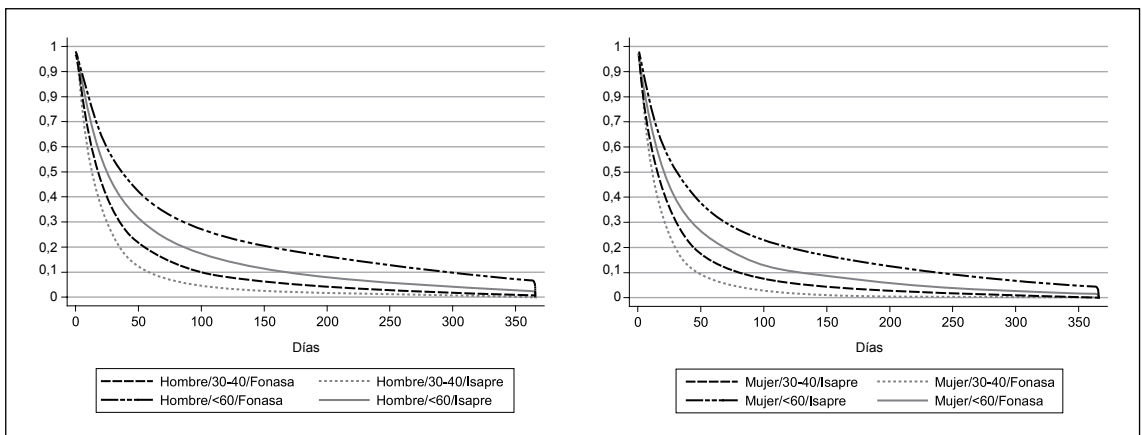

**Figura 3.** Curvas de supervivencia estimada por género, edad y sistema de salud, para la duración de la licencia.

Por otra parte, pertenecer a ISAPRE representa un menor riesgo de poseer licencias médicas más largas que los cotizantes afiliados a FONASA. Complementariamente, la "tasa de ausentismo censurada" propuesta en esta investigación corresponde a 2,4% para los trabajadores suscritos a FONASA y 1,3% a los dependientes de ISAPRE. Estos resultados son coincidentes con los obtenidos por la Dirección de Presupuestos del Gobierno de Chile[4]. En la práctica, esta menor duración del reposo médico puede explicarse por los siguientes motivos: Primero, a las ISAPREs se les permite "descremar" a los afiliados, es decir, pueden conocer *a priori* el mayor riesgo sanitario y quedarse con los mejores clientes, esto es posible por medio de la facultad que tienen de exigir una declaración de salud a cada futuro cotizante[11]. Segundo, en el sistema privado la fiscalización de las licencias es rigurosa y masiva: la ISAPRE puede efectuar una visita domiciliaria a la persona que esté con licencia médica, teniendo por objeto tomar contacto directo y personal con el trabajador a fin de verificar el cumplimiento del reposo indicado en la licencia médica y, de no cumplirse ésta condición, tiene la facultad de rechazar el pago de dicha licencia. Tercero, las empresas aseguradoras aplican una tabla de factores de riesgo para determinar el precio a pagar por cada cotizante, en la cual está definido que, a mayor edad, mayor es el precio del plan de salud; similar discriminación ocurre entre hombres y mujeres. Por lo tanto, esto provoca que las personas mayores y/o mujeres se traspasen al sistema público, en el cual su descuento de salud siempre equivaldrá a 7% de la renta imponible. Entendiendo que, a mayor edad, mayor será el período de licencia médica, nuevamente se percibe que las ISAPREs poseen un método de selección a su favor ("descremar"). Cuarto, otra variable de interés corresponde al copago, es decir la parte monetaria que el cotizante debe sacrificar por poseer una licencia médica; esta restricción afecta a las personas que poseen una tasa de reemplazo menor a 100% debido a que su remuneración es superior al límite máximo imponible. Esto genera un desincentivo para tomar licencias largas, ya que el efecto pecuniario en la renta recibida sería importante. Lo más probable es que este tipo de cotizantes pertenecen al sector sanitario privado, por lo que provocan que el ausentismo promedio de este sector sea menor.

Finalmente, sugerimos generar instancias de debate con respecto a la selección que efectúan legalmente las ISAPREs, específicamente con respecto a la exigencia de la declaración de salud y a la aplicación de la tabla de factores de riesgo. Esto permitiría aliviar la sobre carga de cotizantes del sistema se salud público y posiblemente equilibrar entre ambos sistemas el financiamiento del Subsidio por Incapacidad Laboral.